# On Functional Observability of Nonlinear Systems and the Design of Functional Observers with Assignable Error Dynamics

**Costas Kravaris**

*Artie McFerrin Department of Chemical Engineering, Texas A&M University, College Station, TX, USA
(kravaris@tamu.edu)*

**Abstract**: This paper proposes a novel approach for designing functional observers for nonlinear systems, with linear error dynamics and assignable poles. Sufficient conditions for functional observability are first derived, leading to functional relationships between the Lie derivatives of the output to be estimated and the ones of the measured output. These are directly used in the proposed design of the functional observer. The functional observer is defined in differential input-output form, satisfying an appropriate invariance condition that emerges from the state-space invariance conditions of the literature. A concept of functional observer index is also proposed, to characterize the lowest feasible order of functional observer with pole assignment. Two chemical reactor applications are used to illustrate the proposed approach.

*Keywords*: Functional Observability, Functional Observers, Nonlinear Systems, Exact Linearization

## 1. INTRODUCTION

Functional observers are useful in many applications where a function of the states needs to be estimated, and not the entire state vector. In linear systems, the notion of a functional observer was first defined in Luenberger's pioneer work (Luenberger, 1966 and 1971), where it was proved that it is feasible to construct a functional observer with number of states equal to observability index minus one. In recent years, there has been continued interest in functional observers for linear systems (Darouach, 2000; Korovin et al., 2008 and 2010; Fernando et al., 2010), the goal being to find the smallest possible order of the linear functional observer.

Functional observers for nonlinear systems were defined and formulated in the context of exact linearization (Kravaris, 2016; Kravaris and Venkateswaran, 2021), in the same spirit as in exact linearization of full-state observers (Andrieu and Praly, 2006; Kazantzis and Kravaris, 1998; Kazantzis et al., 2000; Krener and Isidori, 1983; Krener and Respondek, 1985), and have been applied to fault detection and estimation in nonlinear systems (Venkateswaran et al., 2022; Venkateswaran and Kravaris, 2023 and 2024). In the present work, a different approach will be followed. Instead of starting with the requirement of linearity for the error dynamics in state-space form, the starting point will be functional observability, and the functional observer will be derived in input-output form, based on functional relationships between the Lie derivatives of the measured output and the ones of the output to be estimated. The proposed approach will lead to linear error dynamics with assignable poles, for any nonlinear system that is locally functionally observable.

This paper studies unforced nonlinear systems of the form

$$\begin{aligned} \frac{dx}{dt} &= F(x) \\ y &= H(x) \\ z &= q(x) \end{aligned} \qquad (1)$$

where $x \in \mathbb{R}^n$ is the system state, $y \in \mathbb{R}^p$ is the vector of measured outputs, $z \in \mathbb{R}$ is the output to be estimated, $F: \mathbb{R}^n \to \mathbb{R}^n$, $H: \mathbb{R}^n \to \mathbb{R}^p$, $q: \mathbb{R}^n \to \mathbb{R}$ are sufficiently smooth nonlinear functions. The objective is to construct a functional observer which generates an estimate of the output z, driven by the output measurements $y_j, j = 1, \cdots, p$.

Section 2 will start with a brief necessary review of local sate observability and subsequently will define explore key properties of local functional observability for a system of the form (1). It will also define a concept of observability index for nonlinear systems in a way that directly extends Luenberger's definition for linear systems. Section 3 will define functional observer in input-output form, including a direct comparison with the state-space definition of the literature (Kravaris, 2016; Kravaris and Venkateswaran, 2021). The invariance condition of the state-space form will translate to an invariance condition for the input-output form. Section 4 will introduce a notion of functional observer index, to specify the lowest feasible order of functional observer. Sections 5 and 6 will summarize the proposed design approach for a functional observer with linear error dynamics and assignable poles, including two chemical reactor applications. Finally, section 7 will specialize the results to linear systems.

.

## 2. LOCAL STATE OBSERVABILITY AND LOCAL FUNCTIONAL OBSERVABILITY

In this section, local functional observability of nonlinear systems will be defined in a way that is completely analogous to state observability (Montanari et al., 2022), and its properties will be derived in a form that will directly link to the formulation of the functional observer problem of subsequent sections. We will start will a brief necessary review of local state observability, where we will also propose a notion of observability index for nonlinear systems.

State observability is defined as local injectivity of the map $x_0 \mapsto y(t) = H \circ \Phi_F(t; x_0)$, where $\Phi_F(t; x_0)$ denotes the flow of $F(x)$, i.e. the solution of $\frac{dx}{dt} = F(x)$ under the initial condition $x(0) = x_0$.

Definition 1 (Hermann and Krener, 1977; Vidyasagar, 1978): Let $X$ be an open subset of $\mathbb{R}^n$. A nonlinear system of the form (1) will be called state-observable on $X$ if for every $x_1, x_2 \in X$,
$$H \circ \Phi_F(t; x_1) = H \circ \Phi_F(t; x_2) \Rightarrow x_1 = x_2 \tag{2}$$

Consequences of the definition:

Because
$$\frac{d^i}{dt^i}\left[H_j \circ \Phi_F(t; x_0)\right]\bigg|_{t=0} = L_F^i H_j(x_0), \; i = 0, 1, 2, \cdots, \; j = 1, \cdots, p$$
for all $x_0$, the injectivity condition (2) of the definition will be satisfied if it can be established that

$$L_F^i H_j(x_1) = L_F^i H_j(x_2) \text{ for all } i = 0, 1, 2, \cdots, \; j = 1, \cdots, p \tag{3}$$
$$\Rightarrow x_1 = x_2$$

The above says that if a countable number of Lie derivatives being equal at $x_1$ and $x_2$ implies that $x_1 = x_2$, then the system is state-observable. If this property can be established for a finite subset of these Lie derivatives, e.g. up to order (m–1),

$$L_F^i H_j(x_1) = L_F^i H_j(x_2) \text{ for all } i = 0, 1, \cdots, m-1, \; j = 1, \cdots, p \tag{4}$$
$$\Rightarrow x_1 = x_2$$

it will still establish the inference (2) of the definition.

Remark 1: When $F(x)$ and $H(x)$ are real analytic, $H \circ \Phi_F(t; x_0)$ is also real analytic and it is amenable to a local Lie series expansion (Gröbner, 1967):
$$H_j \circ \Phi_F(t; x_0) = \sum_{i=0}^{\infty} L_F^i H_j(x_0) \frac{t^i}{i!}, \; j = 1, \cdots, p \tag{5}$$

Therefore, $H \circ \Phi_F(t; x_1) = H \circ \Phi_F(t; x_2)$ is equivalent to all coefficients of the power series being equal, i.e. $L_F^i H_j(x_1) = L_F^i H_j(x_2), \; i = 0, 1, 2, \cdots, \; j = 1, \cdots, p$, hence inference (3) is also necessary.

Definition 2: The list of functions:
$$\mathcal{O}_{F,H}(x) = \left\{L_F^i H_j(x), \; i = 0, 1, 2, \cdots, \; j = 1, \cdots, p\right\}$$

will be called the observability set of system (1). The list of functions $\mathcal{O}_{F,H,m}(x) = \left\{L_F^i H_j(x), \; i = 0, 1, \cdots, m-1, \; j = 1, \cdots, p\right\}$ will be called the observability set of order (m –1) of system (1).

The notation of Definition 2 will be used throughout this section. Proposition 1 provides sufficient conditions for local state observability; see also Hermann and Krener (1977) and Vidyasagar (1978) that give them in slightly different form.

Proposition 1: (i) *System* (1) *will be locally state-observable if there exists a positive integer* m *such that the Jacobian matrix that consists of the gradients of the elements of* $\mathcal{O}_{F,H,m}(x)$, *i.e. its rows are* $\frac{\partial}{\partial x}\left(L_F^i H_j(x)\right), \; i = 0, 1, \cdots, m-1, \; j = 1, \cdots, p$, *has rank* n.

(ii) *System* (1) *will be locally state-observable if there exists a positive integer* m *such that the mapping* $x \in \mathbb{R}^n \mapsto \mathcal{O}_{F,H,m}(x) \in \mathbb{R}^{p \cdot m}$ *possesses a local left inverse* $x = \varphi\left(L_F^i H_j(x), \; i = 0, 1, \cdots, m-1, \; j = 1, \cdots, p\right)$, *with* $\varphi$ *being a continuously differentiable mapping from a subset of* $\mathbb{R}^{p \cdot m}$ *to* $\mathbb{R}^n$.

(iii) *The above sufficient conditions are equivalent.*

Proof:

-- Condition (i) $\Rightarrow$ n of $L_F^i H_j(x)$ form a locally invertible function from $\mathbb{R}^n$ to $\mathbb{R}^n \Rightarrow$ Condition (ii)

-- Condition (ii) $\Rightarrow$ inference (4) holds $\Rightarrow$ system is locally state observable.

-- Condition (ii) $\Rightarrow$ Jacobian matrix of $\mathcal{O}_{F,H,m}(x)$ is locally a left invertible matrix $\Rightarrow$ Condition (i) □

The following definition provides a direct nonlinear extension of the notion of observability index of a linear system.

Definition 3: *Suppose that system* (1) *satisfies the sufficient condition of state observability of Proposition* 1, *for some positive integer* m. *The smallest positive integer* m *for which it can be satisfied will be called observability index of system* (1) *and will be denoted by* $n_o$.

As an immediate consequence of the definition, the state vector x may be locally represented as a function of $H_j(x), \; L_F H_j(x), \; \ldots, \; L_F^{n_o - 1} H_j(x), \; j = 1, \cdots, p$:
$$x = \varphi\left(H_j(x), \; L_F H_j(x), \; \ldots, \; L_F^{n_o - 1} H_j(x), \; j = 1, \cdots, p\right) \tag{6}$$
with $\varphi$ from a subset of $\mathbb{R}^{p \cdot n_o}$ to $\mathbb{R}^n$.

For unobservable systems, it is well known (see e.g. Vidyasagar, 1978, Theorem 97) that they may be transformed into to a series of an observable system followed by an unobservable system, through an appropriate local coordinate change $\tilde{x} = \begin{bmatrix} \tilde{x}_{ob} \\ \tilde{x}_{unob} \end{bmatrix} = T(x)$, in the form

$$\frac{d\tilde{x}_{ob}}{dt} = \tilde{F}_{ob}(\tilde{x}_{ob})$$
$$\frac{d\tilde{x}_{unob}}{dt} = \tilde{F}_{unob}(\tilde{x}_{ob}, \tilde{x}_{unob}) \quad (7)$$
$$y = \tilde{H}(\tilde{x}_{ob})$$

In this case, it is possible to define observability index $\tilde{n}_o$ for the observable part of the system and obtain a local representation of the observable part of the state vector as

$$\tilde{x}_{ob} = \tilde{\varphi}\left(\tilde{H}_j(\tilde{x}_{ob}), L_{\tilde{F}_{ob}}\tilde{H}_j(\tilde{x}_{ob}), \cdots, L_{\tilde{F}_{ob}}^{\tilde{n}_o-1}\tilde{H}_j(\tilde{x}_{ob}), \; j=1,\cdots,p\right) \quad (8)$$

hence as a function of

$$H_j(x), \; L_F H_j(x), \; \ldots, \; L_F^{\tilde{n}_o-1} H_j(x), \; j=1,\cdots,p.$$

Functional observability refers to the situation where a specific functional can be reconstructed from the measurement signal y, without requiring that the entire state vector x can be reconstructed. In analogy to the definition of state observability, functional observability will be defined as distinguishability of a given functional $z = q(x)$ from the measurement signal $y(t)$: *two different z cannot correspond to the same y(t)*.

Definition 4 (Montanari et al., 2022): *Let $X \subseteq \mathbb{R}^n$ be an open set. A nonlinear system of the form* (1) *will be called functionally observable on $X$ if for every pair of initial states $x_1, x_2 \in X$,*

$$H \circ \Phi_F(t; x_1) = H \circ \Phi_F(t; x_2) \;\Rightarrow\; q(x_1) = q(x_2) \quad (9)$$

Consequences of the definition:

Because
$$\frac{d^i}{dt^i}\left[H_j \circ \Phi_F(t; x_0)\right]\bigg|_{t=0} = L_F^i H_j(x_0), \; i=0,1,2,\cdots, \; j=1,\cdots,p$$

for all $x_0$, inference (9) of the definition will be satisfied if it can be established that

$$L_F^i H_j(x_1) = L_F^i H_j(x_2) \text{ for all } i = 0,1,2,\cdots, \; j=1,\cdots,p \quad (10)$$
$$\Rightarrow q(x_1) = q(x_2)$$

The above says that if a countable number of Lie derivatives being equal at $x_1$ and $x_2$ implies that $q(x_1) = q(x_2)$, then the system is functionally observable. If this property can be established for a finite subset of these Lie derivatives, e.g. up to order (m–1),

$$L_F^i H_j(x_1) = L_F^i H_j(x_2) \text{ for all } i=0,1,\cdots,m-1, \; j=1,\cdots,p \quad (11)$$
$$\Rightarrow q(x_1) = q(x_2)$$

it will still establish the inference (9) of the definition.

Remark 2: Definition 4 and its consequences provide a direct extension of the notion of state observability of Definition 1, where in place of the entire state vector x, a specific functional of x is observable. Every state-observable system is functionally observable for any functional of x.

Proposition 2: *System* (1) *is locally functionally observable if there exists a positive integer m such that $q(x)$ can be locally represented as a function of elements the observability set $\mathcal{O}_{F,H,m}(x)$, i.e.*

$$q(x) = \psi\left(L_F^i H_j(x), \; i=0,1,\cdots,m-1, \; j=1,\cdots,p\right) \quad (12)$$

*with $\psi$ being a mapping from a subset of $\mathbb{R}^{p \cdot m}$ to $\mathbb{R}$.*

Proof: The existence of a representation $q(x) = \psi\left(L_F^i H_j(x), \; i=0,1,\cdots,m-1, \; j=1,\cdots,p\right)$ is equivalent to inference (11), which implies local functional observability. □

Proposition 3: *Suppose that system* (1) *satisfies the sufficient condition of functional observability of Proposition 2, with $\psi$ being smooth. Then:*

(i) *All Lie derivatives of $q(x)$ can be locally represented as functions of elements of the observability set $\mathcal{O}_{F,H}(x)$:*

$$L_F^k q(x) = \psi_k\left(L_F^i H_j(x), \; i=0,1,\cdots,k+m-1, \; j=1,\cdots,p\right), \; k=0,1,2,\cdots \quad (13)$$

*with $\psi_k, \; k=0,1,2,\cdots$ being smooth mappings from a subset of $\mathbb{R}^{p \cdot m}$ to $\mathbb{R}$.*

(ii) *The Jacobian matrix $\mathcal{J}_{F,H,m}(x)$, that consists of the gradients of the elements of $\mathcal{O}_{F,H,m}(x)$, i.e. its rows are $\frac{\partial}{\partial x}\left(L_F^i H_j(x)\right), \; i=0,1,\cdots,m-1, \; j=1,\cdots,p$, has the property that*

$$\text{Rank}\begin{bmatrix}\frac{\partial q}{\partial x}(x) \\ \mathcal{J}_{F,H,m}(x)\end{bmatrix} = \text{Rank}\left[\mathcal{J}_{F,H,m}(x)\right] \quad (14)$$

Proof:

(i) Representation (13) is proved by induction. From (12), it holds for k = 0. Assuming (13) holds for some $k \in \mathbb{N}$, applying the Lie derivative operator $L_F$ on both sides, we conclude that $L_F^{k+1} q(x)$ is a function of the Lie derivatives $L_F^i H_j(x), \; j=1,\cdots,p$, up to order $i = k+m$.

(ii) From (12), it follows that
$$\frac{\partial q}{\partial x}(x) \in \text{span}\left\{\frac{\partial}{\partial x}\left(L_F^i H_j(x)\right), \; i=0,1,\cdots,m-1, \; j=1,\cdots,p\right\},$$
hence the result. □

Remark 3: If system (1) satisfies the sufficient condition of state observability of Proposition 1 and has observability index $n_o$, any functional $q(x)$ can be represented as a function of Lie derivatives of $H(x)$ up to order $n_o - 1$, i.e. representation (12) holds for $m = n_o$. If a system is not state-observable and has been decomposed into observable and unobservable parts according to (7), it will be functionally observable if q(x) can be expressed as a function of the observable part of the state vector $\tilde{x}_{ob}$. In this case, if the observable part of the system has observability index $\tilde{n}_o$, q(x) will be representable as a function of Lie derivatives of H up to order $\tilde{n}_o - 1$.

Remark 4: An alternative way of defining functional observability could be via the requirement that *two different time functions* q(x(t)) *cannot correspond to the same measurement signal* y(t), as follows:

Definition 4′: *Let $X \subseteq \mathbb{R}^n$ be an open set. A nonlinear system of the form* (1) *will be called <u>functionally observable</u> on $X$ if for every pair of initial states* $x_1, x_2 \in X$,
$$H \circ \Phi_F(t; x_1) = H \circ \Phi_F(t; x_2) \Rightarrow q \circ \Phi_F(t; x_1) = q \circ \Phi_F(t; x_2)$$

If one adopts the above alternative definition, the representation (12) of q(x) with ψ smooth will be a sufficient condition for functional observability and the rank property (14) will still hold.

## 3. FUNCTIONAL OBSERVER IN INPUT-OUTPUT FORM

A functional observer is defined in state-space form (Kravaris, 2016; Kravaris and Venkateswaran, 2021) as follows:

Definition 5: *Given a dynamic system of the form* (1), *the system*
$$\frac{d\xi}{dt} = \sigma(\xi, y) \qquad (15)$$
$$\hat{z} = \omega(\xi, y)$$

*where* $\sigma: \mathbb{R}^n \times \mathbb{R}^p \to \mathbb{R}^\nu$, $\omega: \mathbb{R}^n \times \mathbb{R}^p \to \mathbb{R}$ ($\nu < n$) *is called a functional observer for* (1), *if in the series connection*

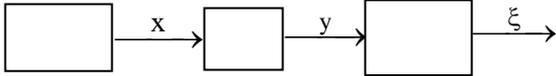

*the overall dynamics*
$$\frac{dx}{dt} = F(x) \qquad (16)$$
$$\frac{d\xi}{dt} = \sigma(\xi, H(x))$$

*possesses an invariant manifold* $\xi = \mathcal{T}(x)$ *with the property that* $q(x) = \omega(\mathcal{T}(x), H(x))$.

The requirement of invariant manifold of (16) translates to
$$\frac{\partial \mathcal{T}}{\partial x}(x)F(x) = \sigma(\mathcal{T}(x), H(x)).$$

Note that, differentiating the output of (15) up to ν times, it is possible to eliminate the states ξ and represent the functional observer in differential input-output form as
$$\frac{d^\nu \hat{z}}{dt^\nu} = \mathcal{G}\left(\hat{z}, \frac{d\hat{z}}{dt}, \cdots, \frac{d^{\nu-1}\hat{z}}{dt^{\nu-1}}, y, \frac{dy}{dt}, \cdots, \frac{d^{\nu-1}y}{dt^{\nu-1}}, \frac{d^\nu y}{dt^\nu}\right) \qquad (17)$$

with $\mathcal{G}$ being a mapping from a subset of $\mathbb{R}^{\nu+(\nu+1)\cdot p}$ to $\mathbb{R}$.

In this work, we will propose a functional observer design method that will lead to an observer equation directly in the form (17). However, before we proceed with the design, we need to specify the required invariance properties of a functional observer in the form (17). Intuitively, we expect that the output of (17) and its time derivatives up to order $\nu - 1$ should be able to accurately track the functional q(x) and its Lie derivatives up to order $\nu - 1$, if correctly initialized:
$$\frac{d^k \hat{z}}{dt^k}(0) = L_F^k q(x(0)) \Rightarrow \frac{d^k \hat{z}}{dt^k}(t) = L_F^k q(x(t)), \quad k = 0, 1, \cdots, \nu-1$$
. (18)

In what follows, we will see how these invariance properties emerge from the state-space definition of the functional observer. In a serial connection of system (17) following system (1), the overall system is described by
$$\frac{dx}{dt} = F(x)$$
$$\frac{d\hat{z}^{(1)}}{dt} = \hat{z}^{(2)}$$
$$\vdots$$
$$\frac{d\hat{z}^{(\nu-1)}}{dt} = \hat{z}^{(\nu)}$$
$$\frac{d\hat{z}^{(\nu)}}{dt} = \mathcal{G}\left(\hat{z}^{(1)}, \ldots, \hat{z}^{(\nu)}, H_j(x), L_F H_j(x), \cdots, L_F^\nu H_j(x), j=1, \cdots, p\right)$$
$$\hat{z} = \hat{z}^{(1)}$$
(19)

System (19) will possess an invariant manifold
$$\begin{bmatrix} \hat{z}^{(1)} \\ \hat{z}^{(2)} \\ \vdots \\ \hat{z}^{(\nu-1)} \\ \hat{z}^{(\nu)} \end{bmatrix} = \mathcal{T}(x) = \begin{bmatrix} q(x) \\ L_F q(x) \\ \vdots \\ L_F^{\nu-1} q(x) \\ L_F^\nu q(x) \end{bmatrix}$$

if and only if the function $\mathcal{G}$ satisfies
$$L_F^\nu q(x) = \mathcal{G}\left(q(x), L_F q(x), \ldots, L_F^{\nu-1} q(x),\right.$$
$$\left. H_j(x), L_F H_j(x), \cdots, L_F^{\nu-1} H_j(x), L_F^\nu H_j(x), j=1, \cdots, p\right)$$

Also notice that $\hat{z} = \hat{z}^{(1)} = q(x)$ on the invariant manifold.
The foregoing considerations lead to the following definition of functional observer in input-output form:

Definition 6: *A dynamic system described by a ν-th order differential equation of the form*
$$\frac{d^\nu \hat{z}}{dt^\nu} = \mathcal{G}\left(\hat{z}, \frac{d\hat{z}}{dt}, \cdots, \frac{d^{\nu-1}\hat{z}}{dt^{\nu-1}}, y_j, \frac{dy_j}{dt}, \cdots, \frac{d^{\nu-1}y_j}{dt^{\nu-1}}, \frac{d^\nu y_j}{dt^\nu}, j=1, \cdots, p\right)$$
(20)

*with $\mathcal{G}$ from $\mathbb{R}^{\nu+(\nu+1)\cdot p}$ to $\mathbb{R}$, is a functional observer for system* (1) *if it satisfies*

$$L_F^\nu q(x) = \mathcal{J}\big(q(x), L_F q(x), \ldots, L_F^{\nu-1} q(x),$$
$$H_j(x), L_F H_j(x), \cdots, L_F^{\nu-1} H_j(x), L_F^\nu H_j(x), j=1,\cdots,p\big) \tag{21}$$

In the presence of initialization errors, additional stability requirements will need to be imposed so that the estimation error asymptotically converges to zero:

$$\lim_{t \to \infty} \left[ \hat{z}(t) - q(x(t)) \right] = 0.$$

In this direction, it is convenient to seek for a functional observer with linear dynamics. In Kravaris and Venkateswaran (2021), a functional observer was sought in the form

$$\frac{d\xi}{dt} = A\xi + \mathcal{B}(y)$$
$$\hat{z} = \omega(C\xi, y) \tag{22}$$

and the stability requirement was satisfied by selecting A to be Hurwitz. Here, we will postulate a linear functional observer of the form

$$\frac{d^\nu \hat{z}}{dt^\nu} + \alpha_1 \frac{d^{\nu-1}\hat{z}}{dt^{\nu-1}} + \cdots + \alpha_{\nu-1}\frac{d\hat{z}}{dt} + \alpha_\nu \hat{z}$$
$$= \mathcal{T}\left(y_j, \frac{dy_j}{dt}, \cdots, \frac{d^{\nu-1}y_j}{dt^{\nu-1}}, \frac{d^\nu y_j}{dt^\nu}, j=1,\cdots,p\right) \tag{23}$$

with $\mathcal{T}$ from $\mathbb{R}^{(\nu+1)\cdot p}$ to $\mathbb{R}$. For this case, invariance condition (21) of Definition 6 takes the form

$$L_F^\nu q(x) + \alpha_1 L_F^{\nu-1} q(x) + \cdots + \alpha_{\nu-1} L_F q(x) + \alpha_\nu q(x)$$
$$= \mathcal{T}(H_j(x), L_F H_j(x), \cdots, L_F^{\nu-1} H_j(x), L_F^\nu H_j(x), j=1,\cdots,p) \tag{24}$$

As long as this condition is satisfied, in the serial connection of system (1) followed by system (23), it will hold that

$$\left(\frac{d^\nu \hat{z}}{dt^\nu} + \alpha_1 \frac{d^{\nu-1}\hat{z}}{dt^{\nu-1}} + \cdots + \alpha_{\nu-1}\frac{d\hat{z}}{dt} + \alpha_\nu \hat{z}\right)$$
$$- \left(L_F^\nu q(x) + \alpha_1 L_F^{\nu-1} q(x) + \cdots + \alpha_{\nu-1} L_F q(x) + \alpha_\nu q(x)\right) = 0$$

or

$$\frac{d^\nu (\hat{z} - q(x))}{dt^\nu} + \alpha_1 \frac{d^{\nu-1}(\hat{z} - q(x))}{dt^{\nu-1}} + \cdots$$
$$+ \alpha_{\nu-1} \frac{d(\hat{z} - q(x))}{dt} + \alpha_\nu (\hat{z} - q(x)) = 0$$

Therefore, the error will asymptotically converge to zero if and only if all the roots of the polynomial

$$\lambda^\nu + \alpha_1 \lambda^{\nu-1} + \cdots + \alpha_{\nu-1}\lambda + \alpha_\nu$$ have negative real parts.

## 4. FUNCTIONAL OBERVER INDEX

From the discussion of the previous section, it is clear that a linear functional observer of order $\nu$ is feasible if we can match a linear combination of the Lie derivatives of the functional up to order $\nu$ with a function of the Lie derivatives of the measurement up to *the same order $\nu$*.

If it so happens that system (1) is state-observable with observability index $n_o$, from (6) it follows that $q(x)$ and its Lie derivatives up to order $n_o - 1$, can be expressed as functions of $H_j(x)$ and its Lie derivatives up to the same order $n_o - 1$:

$$L_F^k q(x) = \psi_k\left(H_j(x), L_F H_j(x), \ldots, L_F^{n_o-1} H_j(x), j=1,\cdots,p\right),$$
$$k = 0, 1, \cdots, n_o - 1 \tag{25}$$

Under these circumstances, a functional observer in input-output form in the sense of Definition 6 can be easily built. For example, if we follow the linear template of equation (23), the defining condition (24) will be satisfied for $\nu = n_o - 1$ and

$$\mathcal{T} = \psi_\nu + \alpha_1 \psi_{\nu-1} + \cdots + \alpha_{\nu-1}\psi_1 + \alpha_\nu \psi_0.$$

The same approach can be followed for any functionally observable system such that the observable part of the system has observability index $\tilde{n}_o$, since a representation of the form (25) will still be valid. Also, it is important to note that a condition of the form (25) could be satisfied for some $\nu$ less than $n_o - 1$, in which case the same construction is applicable and it will lead to a functional observer of lower order.

The foregoing considerations motivate the concept of functional observer index:

<u>Definition 7</u>: *Consider the nonlinear system* (1) *and suppose that there exist a positive integer $\nu$, open sets $X \subseteq \mathbb{R}^n$, $\mathcal{Y} \subseteq \mathbb{R}^{(\nu+1)\cdot p}$ and functions $\psi_k : \mathcal{Y} \to \mathbb{R}$, $k = 0, 1, \cdots, \nu$, such that*

$$L_F^k q(x) = \psi_k\left(H_j(x), L_F H_j(x), \cdots, L_F^\nu H_j(x), j=1,\cdots,p\right), k=0,1,\cdots,\nu \ \forall x \in X \tag{26}$$

*The smallest integer $\nu$ for which the above holds, is called <u>functional observer index</u> for the nonlinear system* (1).

<u>Immediate consequences of the definition</u>:

a) *If a system of the form* (1) *satisfies condition* (26) *of Definition 7 for some positive integer $\nu$, it is functionally observable.*

b) *Suppose that system* (1) *satisfies the sufficient condition of state observability of Proposition* 1 *and has observability index $n_o$. Then it possesses a functional observer index $\nu$, and it holds that $\nu \leq n_o - 1$.*

c) *Suppose that system* (1) *has been decomposed into observable and unobservable parts in the form* (7) *and that the observable part satisfies the sufficient condition of Proposition* 1 *and has observability index $\tilde{n}_o$. Then it possesses a functional observer index $\nu$, and it holds that $\nu \leq \tilde{n}_o - 1$.*

## 5. FUNCTIONAL OBSERVER DESIGN WITH ASSIGNABLE ERROR DYNAMICS

The derivations of the previous sections immediately lead to the following Proposition, which is the main result of the present paper:

Proposition 4: *Consider the nonlinear system* (1) *and suppose that there exist a positive integer v, open sets* $X \subseteq \mathbb{R}^n$, $Y \subseteq \mathbb{R}^{(v+1)\cdot p}$ *and functions* $\psi_k : Y \to \mathbb{R}$, $k = 0,1,\cdots,v$, *such that*

$$L_F^k q(x) = \psi_k\left(H_j(x), L_F H_j(x), \cdots, L_F^v H_j(x), j=1,\cdots,p\right), \; k=0,1,\cdots,v \quad \forall x \in X \tag{26}$$

*Also, let* $\lambda^v + \alpha_1 \lambda^{v-1} + \cdots + \alpha_{v-1}\lambda + \alpha_v$ *be a given polynomial whose roots all have negative real parts.*
*Then the dynamic system*

$$\frac{d^v \hat{z}}{dt^v} + \alpha_1 \frac{d^{v-1}\hat{z}}{dt^{v-1}} + \cdots + \alpha_{v-1}\frac{d\hat{z}}{dt} + \alpha_v \hat{z}$$
$$= \mathcal{T}\left(y_j, \frac{dy_j}{dt}, \cdots, \frac{d^{v-1}y_j}{dt^{v-1}}, \frac{d^v y_j}{dt^v}, j=1,\cdots,p\right) \tag{23}$$

*where*

$$\mathcal{T} = \psi_v + \alpha_1 \psi_{v-1} + \cdots + \alpha_{v-1}\psi_1 + \alpha_v \psi_0 \tag{27}$$

*is a functional observer for system* (1)*, and has the property that the estimation error* $[\hat{z}(t) - q(x(t))]$ *converges asymptotically to zero, following*

$$\frac{d^v(\hat{z}-q(x))}{dt^v} + \alpha_1 \frac{d^{v-1}(\hat{z}-q(x))}{dt^{v-1}} + \cdots$$
$$+ \alpha_{v-1}\frac{d(\hat{z}-q(x))}{dt} + \alpha_v(\hat{z}-q(x)) = 0 \tag{28}$$

Some comments:

a) The derived design equations lead to a v-th order functional observer, for <u>any</u> positive v that can satisfy (26). However, from the point of view of ease of implementation, it is desirable to have the observer order as low as possible. Consequently, in practical applications, v is expected to be chosen to be the functional observer index.

b) Because the polynomial $\lambda^v + \alpha_1 \lambda^{v-1} + \cdots + \alpha_{v-1}\lambda + \alpha_v$, whose roots govern the error dynamics (28) may be arbitrarily picked, this means that the design enables assignment of the rate of decay of the error.

c) The derived observer is in differential input-output form, following a v-th order differential equation that relates the input (measurement signal) to the output (estimate). The input-output description is proper in the sense that the order of input derivatives does not exceed the order of output derivatives. Consequently, the observer can be appropriately discretized in a causal form and simulated. However, in is only in special types of nonlinear functions $\psi_0, \psi_1, \cdots, \psi_v$ that a state-space representation of the form (22) will be possible (see specific examples in the applications section that follows).

The conclusion from this section is that *local functional observability enables the design of a functional observer with assignable error dynamics*.

## 6. CHEMICAL REACTOR APPLICATIONS

*6.1 Isothermal batch chemical reactor*

Consider an isothermal batch reactor where consecutive irreversible chemical reactions A → B → C → D take place. The first and third reactions have first order kinetics, whereas the second reaction has second-order kinetics. The reactor dynamics can be modelled through standard component mass balances for species A, B, C, assuming constant reactor volume, as follows:

$$\frac{dc_A}{dt} = -k_1 c_A$$
$$\frac{dc_B}{dt} = k_1 c_A - k_2 c_B^2$$
$$\frac{dc_C}{dt} = k_2 c_B^2 - k_3 c_C$$

where $c_A$, $c_B$, $c_C$ are the concentrations of species A, B, C in the reactor and $k_1, k_2, k_3$ are the reaction rate constants. The second sate is measured, whereas the first state needs to be estimated:

$$y = c_B$$
$$z = c_A$$

From the second equation, we can calculate

$$z = c_A = \frac{1}{k_1}\left(\frac{dc_B}{dt} + k_2 c_B^2\right) = \frac{1}{k_1}\left(\frac{dy}{dt} + k_2 y^2\right),$$

whereas from the first equation,

$$\frac{dz}{dt} = \frac{dc_A}{dt} = -k_1 c_A = -\left(\frac{dc_B}{dt} + k_2 c_B^2\right) = -\left(\frac{dy}{dt} + k_2 y^2\right)$$

So, it is possible to express the unmeasured output and its time derivative in terms of the measured output and its derivative. This means that the functional observer index is v = 1. The corresponding functions $\psi_0$ and $\psi_1$ are given by:

$$\psi_0\left(y, \frac{dy}{dt}\right) = \frac{1}{k_1}\left(\frac{dy}{dt} + k_2 y^2\right)$$
$$\psi_1\left(y, \frac{dy}{dt}\right) = -\left(\frac{dy}{dt} + k_2 y^2\right)$$

and the resulting functional observer is as follows (here we have set $\alpha_1 = -\lambda$, where λ is the eigenvalue of the error dynamics):

$$\frac{d\hat{z}}{dt} - \lambda \hat{z} = -\left(1 + \frac{\lambda}{k_1}\right)\frac{dy}{dt} - \left(1 + \frac{\lambda}{k_1}\right)k_2 y^2$$

The above observer is in input-output form. Alternatively, it may be represented in state space form as follows:

$$\begin{cases} \dfrac{d\xi}{dt} = \lambda \xi - \left(1 + \dfrac{\lambda}{k_1}\right)\left(\lambda y + k_2 y^2\right) \\ \hat{z} = \xi - \left(1 + \dfrac{\lambda}{k_1}\right) y \end{cases}$$

## 6.2 Nonisothermal continuous chemical reactor

Consider a non-isothermal Continuous Stirred Tank Reactor (CSTR) where an irreversible exothermic chemical reaction A → B with first order kinetics takes place. The reactor is cooled through a cooling jacket. The reactor dynamics can be modelled through standard component mass balances and energy balances, assuming constant volume and constant thermophysical properties, as follows:

$$\frac{dc_A}{dt} = \frac{F}{V}(c_{A_{in}} - c_A) - k(\theta)c_A$$

$$\frac{d\theta}{dt} = \frac{F}{V}(\theta_{in} - \theta) + \frac{(-\Delta H)_R}{\rho c_p}k(\theta)c_A - \frac{UA}{\rho c_p V}(\theta - \theta_J)$$

$$\frac{d\theta_J}{dt} = \frac{F_J}{V_J}(\theta_{J_{in}} - \theta_J) + \frac{UA}{\rho_J c_{p_J} V_J}(\theta - \theta_J)$$

where $c_A$ is the concentration of species A in the reacting mixture, $\theta$ and $\theta_J$ are the temperatures of the reacting mixture and the jacket fluid respectively; these are the system states. The function $k(\theta) = k_0 e^{-\frac{E}{R\theta}}$ is the Arrhenius correlation that expresses the reaction rate constant as a function of temperature. The rest of the symbols represent constant parameters: $c_{Ain}$ is the feed concentration of species A, F and $F_J$ are the feed and coolant flowrates respectively, V and $V_J$ are the reactor volume and cooling jacket volume respectively, $(-\Delta H)_R$ is the heat of reaction, $\rho, c_p$ and $\rho_J, c_{p_J}$ are the densities and heat capacities of the reactor contents and cooling fluid respectively, U and A are the overall heat transfer coefficient and heat transfer area respectively.

The second and third states are measured: $\begin{aligned} y_1 &= \theta \\ y_2 &= \theta_J \end{aligned}$

whereas the first state needs to be estimated: $z = c_A$

From the 2$^{nd}$ equation,

$$c_A = \frac{\rho c_p V \frac{d\theta}{dt} - F\rho c_p(\theta_{in} - \theta) + UA(\theta - \theta_J)}{(-\Delta H)_R V k(\theta)},$$

then from the 1$^{st}$ equation,

$$\frac{dc_A}{dt} = \frac{F}{V}c_{A_{in}} - \left(\frac{F}{V} + k(\theta)\right)c_A$$

$$= \frac{F}{V}c_{A_{in}} - \frac{\rho c_p V \frac{d\theta}{dt} - F\rho c_p(\theta_{in} - \theta) + UA(\theta - \theta_J)}{(-\Delta H)_R V}\left(\frac{F}{Vk(\theta)} + 1\right)$$

So, it is possible to express the unmeasured output and its time derivative in terms of the measured outputs and their derivatives. This means that the functional observer index is $\nu = 1$. The corresponding functions $\psi_0$ and $\psi_1$ are given by:

$$\psi_0\left(y, \frac{dy}{dt}\right) = \frac{\rho c_p}{(-\Delta H)_R} \cdot \frac{1}{k(\theta)}\frac{d\theta}{dt} - \frac{F\rho c_p(\theta_{in} - \theta) - UA(\theta - \theta_J)}{(-\Delta H)_R V k(\theta)}$$

$$\psi_1\left(y, \frac{dy}{dt}\right) = -\frac{\rho c_p}{(-\Delta H)_R}\left(\frac{F}{Vk(\theta)} + 1\right)\frac{d\theta}{dt}$$

$$+ \frac{F\rho c_p(\theta_{in} - \theta) - UA(\theta - \theta_J)}{(-\Delta H)_R V}\left(\frac{F}{Vk(\theta)} + 1\right) + \frac{F}{V}c_{A_{in}}$$

and the resulting functional observer follows the equation (here we have set $\alpha_1 = -\lambda$, where $\lambda$ is the eigenvalue of the error dynamics):

$$\frac{d\hat{z}}{dt} - \lambda\hat{z} = -\frac{\rho c_p}{(-\Delta H)_R}\left(\frac{\frac{F}{V} + \lambda}{k(\theta)} + 1\right)\frac{d\theta}{dt}$$

$$+ \frac{F\rho c_p(\theta_{in} - \theta) - UA(\theta - \theta_J)}{(-\Delta H)_R V}\left(\frac{\frac{F}{V} + \lambda}{k(\theta)} + 1\right) + \frac{F}{V}c_{A_{in}}$$

The above observer is in input-output form. Alternatively, it may be represented in state space form as follows:

$$\frac{d\xi}{dt} = \lambda\xi - \lambda\frac{\rho c_p}{(-\Delta H)_R}\left(\theta + \left(\frac{F}{V} + \lambda\right)\int_0^\theta \frac{d\zeta}{k(\zeta)}\right)$$

$$+ \frac{F\rho c_p(\theta_{in} - \theta) - UA(\theta - \theta_J)}{(-\Delta H)_R V}\left(\frac{\frac{F}{V} + \lambda}{k(\theta)} + 1\right) + \frac{F}{V}c_{A_{in}}$$

$$\hat{z} = \xi - \frac{\rho c_p}{(-\Delta H)_R}\left(\theta + \left(\frac{F}{V} + \lambda\right)\int_0^\theta \frac{d\zeta}{k(\zeta)}\right)$$

## 7. FUNCTIONAL OBSERVER DESIGN FOR LINEAR SYSTEMS

The design method developed in the previous sections can now be specialized to linear systems. Consider

$$\begin{aligned} \frac{dx}{dt} &= Fx \\ y &= Hx \\ z &= qx \end{aligned} \quad (29)$$

with F, H, q being $n \times n$, $p \times n$, $1 \times n$ matrices respectively, where y is the measured output and z is the functional to be estimated.

Suppose that for some $\nu \in \mathbb{N}$,

$q, qF, \cdots, qF^{\nu-1}, qF^\nu \in \text{span}\{H_j, H_jF, \ldots, H_jF^\nu, j = 1, \cdots, p\}$,

i.e. there is a $(\nu+1) \times p(\nu+1)$ matrix M such that

$$\begin{bmatrix} q \\ qF \\ \vdots \\ qF^{\nu-1} \\ qF^\nu \end{bmatrix} = M \begin{bmatrix} H \\ HF \\ \vdots \\ HF^{\nu-1} \\ HF^\nu \end{bmatrix}.$$

Let $\lambda^\nu + \alpha_1 \lambda^{\nu-1} + \cdots + \alpha_{\nu-1}\lambda + \alpha_\nu$ be a given polynomial whose roots all have negative real parts.
Then

$$qF^\nu + \alpha_1 qF^{\nu-1} + \cdots + \alpha_{\nu-1} qF + \alpha_\nu q$$

$$= [\alpha_\nu \;\; \alpha_{\nu-1} \;\; \cdots \;\; \alpha_1 \;\; 1] M \begin{bmatrix} H \\ HF \\ \vdots \\ HF^{\nu-1} \\ HF^\nu \end{bmatrix}$$

and defining $[\beta_\nu \;\; \beta_{\nu-1} \;\; \cdots \;\; \beta_1 \;\; \beta_0] = [\alpha_\nu \;\; \alpha_{\nu-1} \;\; \cdots \;\; \alpha_1 \;\; 1] M$, we can write

$$qF^\nu + \alpha_1 qF^{\nu-1} + \cdots + \alpha_{\nu-1} qF + \alpha_\nu q$$
$$= \beta_0 HF^\nu + \beta_1 HF^{\nu-1} + \cdots + \beta_{\nu-1} HF + \beta_\nu H$$

Thus, the system

$$\frac{d^\nu \hat{z}}{dt^\nu} + \alpha_1 \frac{d^{\nu-1}\hat{z}}{dt^{\nu-1}} + \cdots + \alpha_{\nu-1}\frac{d\hat{z}}{dt} + \alpha_\nu \hat{z}$$
$$= \beta_0 \frac{d^\nu y}{dt^\nu} + \beta_1 \frac{d^{\nu-1} y}{dt^{\nu-1}} + \cdots + \beta_{\nu-1}\frac{dy}{dt} + \beta_\nu y \quad (30)$$

is a functional observer of order $\nu$ in input-output form, whose error dynamics are governed by the roots of the polynomial $\lambda^\nu + \alpha_1 \lambda^{\nu-1} + \cdots + \alpha_{\nu-1}\lambda + \alpha_\nu$. The observer's input-output description (30) can be converted in state-space form:

$$\frac{d\xi}{dt} = A\xi + By$$
$$\hat{z} = C\xi + Dy$$

with (C, A) in observer canonical form, as follows:

$$A = \begin{bmatrix} 0 & 0 & \cdots & 0 & -\alpha_\nu \\ 1 & 0 & \cdots & 0 & -\alpha_{\nu-1} \\ 0 & 1 & \cdots & 0 & -\alpha_{\nu-2} \\ \vdots & \vdots & & \vdots & \vdots \\ 0 & 0 & \cdots & 1 & -\alpha_1 \end{bmatrix}, \;\; B = \begin{bmatrix} \beta_\nu - \alpha_\nu \beta_0 \\ \beta_{\nu-1} - \alpha_{\nu-1}\beta_0 \\ \beta_{\nu-2} - \alpha_{\nu-2}\beta_0 \\ \vdots \\ \beta_1 - \alpha_1 \beta_0 \end{bmatrix}$$
$$C = [\; 0 \;\; 0 \;\; \cdots \;\; 0 \;\; 1 \;], \;\; D = \beta_0$$

## 8. CONCLUSIONS

This paper has developed a novel approach for designing functional observers with linear error dynamics and pole placement. Design is based on local functional observability analysis, utilizing relationships between the Lie derivatives of the output to be estimated and the measured output. The method is illustrated through chemical reactor applications.

## ACKNOWLEDGEMENTS

Financial support from the National Science Foundation through the grant CBET-2133810 is gratefully acknowledged.